\documentstyle[prl,aps,epsfig]{revtex}
\begin{document}
\draft 
\twocolumn[\hsize\textwidth\columnwidth\hsize\csname@twocolumnfalse%
\endcsname
\title {
 Electroproduction of the $\Delta(1232)$ resonance at  high momentum transfer.} 
\author {
V.~V.~Frolov$^{1\ast}$, 
G.S.~Adams$^1$, 
A.~Ahmidouch$^5$, 
C.~S.~Armstrong$^3$, 
K.~Assamagan$^4$, 
S.~Avery$^4$, 
O.~K.~Baker$^{4,7}$, 
P.~Bosted$^2$, 
V.~Burkert$^7$, 
R.~Carlini$^7$, 
R.~M.~Davidson$^1$, 
J.~Dunne$^7$, 
T.~Eden$^4$, 
R.~Ent$^7$, 
D.~Gaskell$^4$, 
P.~Gu\`eye$^4$, 
W.~Hinton$^4$, 
C.~Keppel$^{4,7}$, 
W.~Kim$^6$, 
M.~Klusman$^1$, 
D.~Koltenuk$^8$, 
D.~Mack$^7$, 
R.~Madey$^{4,5}$, 
D.~Meekins$^3$, 
R.~Minehart$^9$
J.~Mitchell$^7$, 
H.~Mkrtchyan$^{10}$, 
N.~C.~Mukhopadhyay$^1$, 
J.~Napolitano$^1$, 
G.~Niculescu$^4$, 
I.~Niculescu$^4$, 
M.~Nozar$^1$, 
J.~W.~Price$^{1+}$, 
P.~Stoler$^1$, 
V.~Tadevosian$^{10}$, 
L.~Tang$^{4,7}$, 
M.~Witkowski$^1$, 
S.~Wood$^7$.} 
\address{\it $^1$Physics Department, Rensselaer Polytechnic Institute, Troy NY 12180 } 
\address{\it $^2$Physics Department, American University, Washington, D.C. 20016 } 
\address{\it $^3$Department of Physics, College of William \& Mary, Williamsburg, VA 23187 } 
\address{\it $^4$Physics Department, Hampton University, Hampton, VA 23668 } 
\address{\it $^5$Physics Department, Kent State University, Kent OH 44242 } 
\address{\it $^6$Physics Department, Kyungpook National University, Taegu, South Korea } 
\address{\it $^7$Thomas Jefferson National Accelerator Facility, Newport News VA 23606 } 
\address{\it $^8$Physics Department, University of Pennsylvania, Philadelphia PA 19104 } 
\address{\it $^9$Physics Department, University of Virginia, Charlottesville, VA 22903 } 
\address{\it $^{10}$Yerevan Physics Institute, Yerevan, Armenia } 
\date{\today}
\maketitle
\begin{abstract}
We studied the electroproduction of the $\Delta(1232)$ resonance via the reaction 
$p(e,e^\prime p)\pi^0$ at  four-momentum transfers $Q^2$ = 2.8 and 4.0 GeV$^2$.
This is the highest $Q^2$ for which exclusive resonance electroproduction has ever been 
observed. Decay angular distributions for  $\Delta \to p\pi^0$ were measured over a wide 
range of barycentric energies covering the resonance. The $N-\Delta$ transition form factor
$G^\ast_M$ and ratios of resonant multipoles $E_{1+}/M_{1+}$ and $S_{1+}/M_{1+}$ were extracted from the 
decay angular distributions. These ratios remain small, indicating that perturbative QCD 
is not applicable for this reaction at these momentum transfers.
\end{abstract}
]
An important concern in hadron physics is the determination of  the appropriate degrees of freedom to 
describe exclusive reactions in the experimentally accessible range of momentum transfers.
It is widely agreed that perturbative QCD (pQCD) should apply at sufficiently high
momentum transfer; however, there is no general agreement about
how high the momentum transfer must be.
Many exclusive reactions exhibit scaling behavior which has been interpreted by some authors 
\cite{brod} as the onset of pQCD. Others 
\cite{isg} have argued pQCD would apply only at much higher $Q^2$, and that the observed scaling
can be  explained by contributions 
of the soft Feynman mechanism.

The evolution from  soft non-perturbative physics towards hard pQCD can be studied
using the excitation of the $\Delta(1232)$ resonance. 
In addition to the $Q^2$ dependence
of the transition form factor, $G^\ast_M$, pQCD makes definite predictions about the 
relative contributions of the magnetic dipole, $M_{1+}$, electric quadrupole, $E_{1+}$, 
and Coulomb quadrupole, $S_{1+}$, multipole amplitudes. At low $Q^2$, according to the  
quark model  the $N-\Delta$ transition is due primarily  to a single quark spin flip
so that $M_{1+}$ would dominate, and
the contributions of $E_{1+}$ and $S_{1+}$ would be  very small \cite{bemo}.
 Near $Q^2=0$, recent experiments confirm this prediction: 
$R_{EM} \equiv E_{1+}/M_{1+}  \sim -0.03$  \cite{photo}  and $R_{SM} 
\equiv S_{1+}/M_{1+} \sim -0.11$\cite{elec}, where $E_{1+}$ and $M_{1+}$ are evaluated
at the resonance position.
 At high $Q^2$, according to valence pQCD\cite{brod}, only helicity-conserving amplitudes should 
contribute, leading to the predictions $R_{EM}=+1$ and 
$R_{SM}\rightarrow~constant$.
An evaluation  of earlier data from DESY \cite{haid} at $Q^2$ = 3.2~GeV$^2$ suggests
 that $R_{EM}$ is small, but with large errors.

To address these issues, we  measured  the differential cross section  of the neutral pion
decay channel in electroproduction of the $\Delta(1232)$ at  $Q^2$ = 2.8
and 4.0 GeV$^2$. This experiment was performed with the 100$\%$ duty 
factor electron beam in Hall C at the Thomas Jefferson National Accelerator Facility.
The availability of a multi-GeV beam having high duty factor, high luminosity and
excellent beam energy resolution allowed us for the first time to measure this
exclusive reaction with high statistical precision in this  range of $Q^2$.  

Electron  beams having energies  3.2 and 4.0 GeV, with 
currents  100 and 80~$\mu$A at $Q^2 = 2.8$~and~4.0~GeV$^2$, respectively,
were incident on a liquid hydrogen target of thickness 4  cm.
 The beam current was measured to 1$\%$ accuracy by two resonant cavities and a parametric 
current transformer.

At these $Q^2$ the protons emerge in a narrow cone around the momentum transfer
vector $\vec{q}$. 
Thus, a significant fraction of the full c.m.\ solid angle was obtained in the decay of the
$\Delta \to p\pi^0$. Scattered electrons were detected in the Short Orbit Spectrometer (SOS)
and protons were detected in the High Momentum Spectrometer (HMS).
The SOS central momentum and angle were fixed throughout the data taking at each $Q^2$ point.
 The HMS central momentum and angle were varied to cover the proton's full momentum 
range and most of its decay angle cone. There were a total of 45  kinematic settings at $Q^2$ = 2.8 GeV$^2$  and 28 settings at 4.0 GeV$^2$. The momentum and angular acceptances of adjacent
settings were overlapped for redundancy.

 Electrons were identified utilizing  a  gas threshold
\v{C}erenkov counter and a lead glass shower counter. 
 Protons were identified from the coincidence time between electron
and hadron arms, and by  single arm time-of-flight.

\begin{figure}[tbp]
\begin{center}
\epsfig{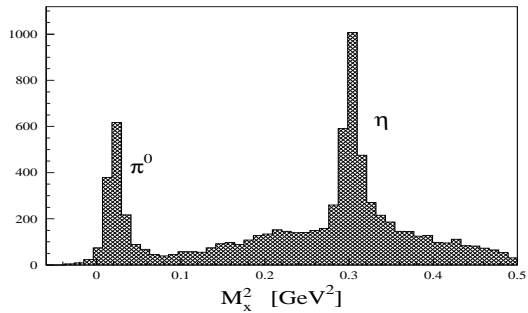}
\end{center}
\caption{Example of a  missing mass distribution for the reaction $p(e,e'p)X$. 
The data are from one experimental setting at $Q^2 = 2.8$ GeV$^2$ ($1.1<W<1.7$ GeV).}
\label{mm}
\end{figure}

The experiment is described in detail in refs. \cite{vvf,csa}.
The single pion final state was selected from the missing mass  of the reaction
$e+p\rightarrow e^\prime + p^\prime + X$.
 A typical experimental missing mass distribution 
is shown in Fig.~\ref{mm}. The $\eta$ production data shown in the figure are dealt
with  elsewhere \cite{csa}.

A brief outline of the analysis follows. For the $p\pi^0$ system, we calculated the invariant mass $W$, the polar  center-of-mass decay 
angle $\cos\theta^{cm}_\pi$, and the out-of-plane angle  $\phi^{cm}_\pi$. Most of
the background from the elastic radiative tail was removed by accepting only those events
with ${\mid \phi^{cm}_\pi \mid} > 8^\circ$ and which passed the missing mass $\pi^0$ cut. The 
remaining contribution ($\sim$ 5$\%$) was modeled and subtracted by a Monte Carlo 
simulation.  The accidental background ($\sim 1 \%$) and
the background associated with events rescattered on the spectrometer collimator, vacuum
chamber, and magnet ($\sim 5 \%$) were obtained from experimental data which were outside
the timing and track reconstruction cuts, and subtracted on a bin-by-bin basis. 
The data were  corrected also for tracking inefficiency ($\sim 9\%$), trigger inefficiency 
($< 1\%$), dead time ($\sim 2.5 \%$), nuclear absorption ($\sim 3.5 \%$), 
local heating effects on target density  
($\sim 4 \%$), and target wall contributions ($\sim 1.5 \%$). 
The acceptance and radiative corrections were obtained from  
a Monte Carlo simulation of the entire experiment. The product of the target 
thickness and beam current was monitored throughout the experiment by the electron single arm 
rate. The ratio of the number of SOS counts to the beam charge was stable to $1\%$ from run 
to run. To check the absolute normalization,  elastic scattering measurements were performed. 
The result is consistent with world data \cite{bosted} to within $2\%$.

Assuming the one-photon-exchange approximation, the differential cross section of single
pion electroproduction is related to the center-of-mass differential cross section for pion 
production by virtual photons, ${d\sigma}/{d\Omega_\pi^{cm}}$, as follows:
\begin{eqnarray}
\label{lt_eq}
\frac{d\sigma}{dW dQ^2 d\Omega_\pi^{cm}} 
 = \Gamma_v \frac{d\sigma}{d\Omega_\pi^{cm}}, 
\end{eqnarray}
where $\Gamma_v$ is the virtual photon flux factor. The center-of-mass 
angles,  $\theta^{cm}_\pi$ and $\phi^{cm}_\pi$,  were reconstructed from the detected
proton laboratory angles. Because at these  $Q^2$ the protons emerge in a 
narrow cone in the laboratory,  accurate measurements of the proton and electron
momenta and angles are necessary to produce good angular resolution in the c.m.\
The resolutions obtained were typically $\sigma_W = 15$ MeV, $\sigma_{Q^2} =
 0.006$ GeV$^2$, $\sigma_{cos\theta^{cm}} = 0.03$ and
$\sigma_{\phi^{cm}} = 3^\circ$.

\begin{figure}[tbp]
\begin{center}
\epsfig{file=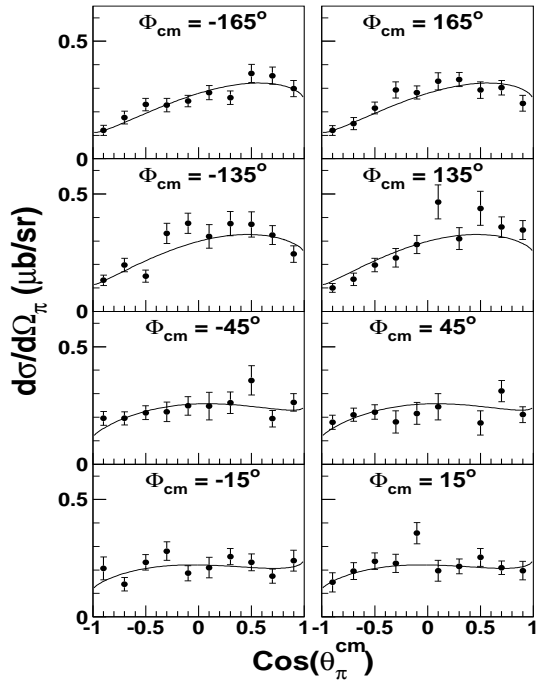,width=3.0in,height=4.0in}
\end{center}
\caption{Examples of the angular distributions for the data at $W=1.235$~GeV and 
$Q^2=4$~GeV$^2$. The curves are the result of an empirical  model-independent fit to the data
at this $W$ value, assuming that only $S$ and $P$ waves contribute and $M_{1+}$ dominance.}
\label{xsn}
\end{figure}

The events with invariant mass $1.1<W<1.4$~GeV were binned with $\Delta W = 30$~MeV, 
$\Delta \cos\theta^{cm}_\pi = 0.2$, and $\Delta \phi^{cm}_\pi = 30^\circ$. The $Q^2$ bin size
($\approx$  0.5 GeV$^2$), was determined by the apparatus acceptance. 
Measurements of
${d\sigma}/{d\Omega_\pi^{cm}}$ were obtained, with the Hand convention
for $\Gamma_v$,  at 751 intervals of 
$\cos \theta^{cm}_\pi$, $\phi^{cm}_\pi$, and $W$ at $Q^2 = 2.8$ and 867 intervals at $Q^2$ =
 4.0~GeV$^2$.
Examples of the angular distribution are shown in Fig.~\ref{xsn}. Errors shown in the
figure are 
statistical only; they do not include  an estimated systematic uncertainty 
in the overall normalization of 5$\%$.

We extracted information about the contributing multipoles by two methods. One
method consisted of making model-independent {\em empirical} multipole fits to the 
angular distributions independently
at each $W$, assuming $M_{1+}$ dominance at the resonance pole, and only
$S$ and $P$ wave contributions, as, for example in \cite{alder}. The quality of the
fits is indicated in Fig.~\ref{xsn}.  Over the entire dataset  
$\chi^2 \approx $ 1.36 per degree of freedom at  $Q^2$ = 2.8  GeV$^2$ and 1.21 per degree of freedom at $Q^2$ = 4.0 GeV$^2$. The resulting extracted
amplitudes confirm the $M_{1+}$ dominance at the resonance.
 In particular at $W=1.235$ GeV,
 at  Q$^2$ = 2.8 GeV$^2$, $R_{EM} = - 0.01 \pm 0.01$  and  $R_{SM} = - 0.06 \pm 0.01$, and at  Q$^2$ = 4.0 GeV$^2$,  $R_{EM} = - 0.02 \pm 0.01$ 
and  $R_{SM} = - 0.11 \pm 0.01$.
This result shows clearly that at the resonance position
$R_{EM}$ is very small and $R_{SM}$ is moderately small and negative.

\begin{figure}[t]
\begin{center}

\epsfig{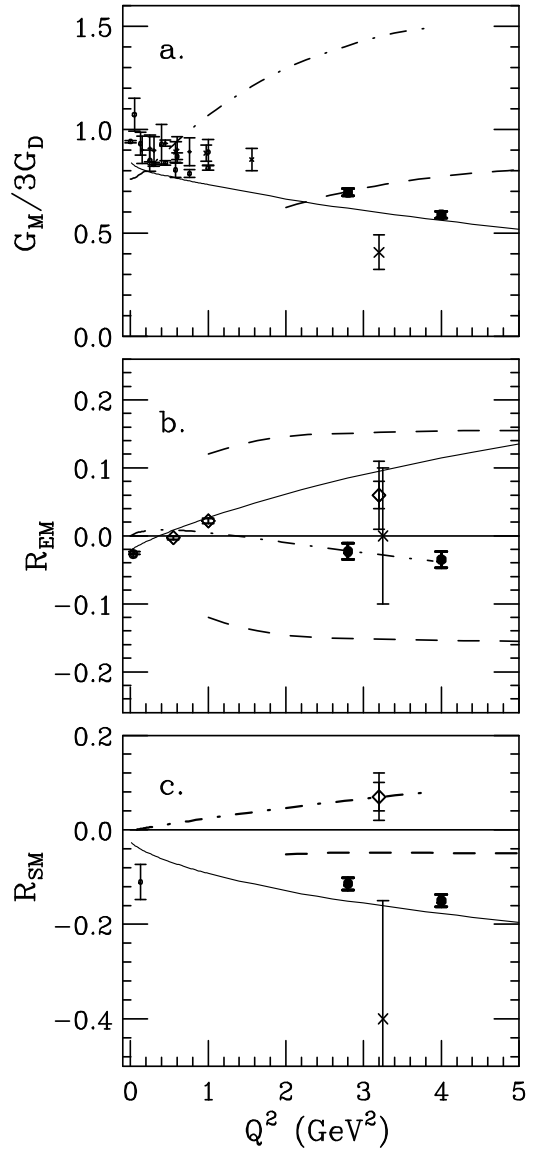}
\end{center}
\caption[h]{Transition form factor $G_M^\ast$ and ratios $R_{EM}$ and $R_{SM}$.
The data points at $Q^2$ =  2.8 and 4.0 ~GeV$^2$, denoted by $\bullet$ are extractions of $G_M^\ast$, $R_{EM}$ and $R_{SM}$ from the present experimental
data, as discussed in the text. 

(a.)  $G_M^\ast$: The data for $Q^2 < 2$ ~GeV$^2$ are evaluations of previous
exclusive data given by \protect\cite{sto,dav,kep,burk}. The datum denoted
 $\times$ at  $Q^2 = 3.2 $ ~GeV$^2$ is  the evaluation of  DESY data
\protect\cite{haid}  by  \protect\cite{dav}.
 (b.)  $R_{EM}$: The data at $Q^2 =0$ are from  \protect\cite{photo}.
 (c.)  $R_{SM}$: the datum at $Q^2 = 0.13$ ~GeV$^2$  is from  \protect\cite{elec}. 
 For  $R_{EM}$ and  $R_{SM}$ the
data denoted by diamonds ($\diamond$) at  $Q^2$ =0.55, 1.0, and 3.2  ~GeV$^2$ 
are evaluations of DESY data \protect\cite{haid} from \protect\cite{burk}, and those denoted
by $\times$ at  $Q^2 = 3.2 $ ~GeV$^2$ are the evaluation of the DESY data by  
\protect\cite{dav}. 

The solid and dot-dash  curves are due to the CQM light-front calculations
with  \protect\cite{card}, and without  \protect\cite{cap},
intrinsic  constituent quark form-factors respectively.  The dashed curves are  the 
results of a pQCD sum rule calculation  \protect\cite{bel}, where the two dashed curves
in $R_{EM}$ are upper and lower limits. 
}
\label{results}
\end{figure}

A more sophisticated extraction of the amplitudes was performed using an effective 
Lagrangian (EL) theoretical basis \cite{dmw}. 
In this approach, the $s$- and $u$-channel
contributions of the nucleon and $\Delta$ are taken into account, along with the $t-$channel vector meson 
contributions. The full $\gamma N \Delta$ and $\pi N \Delta$ vertices are retained,
and the characteristic ``off-shell'' parameters for the $\Delta$ and 
the electromagnetic gauge couplings are fitted to the data at 
each $Q^2$\cite{q2dep}, maintaining unitarity.
Similar fits to existing data at lower $Q^2$ have been performed by other authors
with results close to what we give here \cite{sato}.  The  EL extracted values of
$G^\ast_M/3G_D$, $R_{EM}$,  and  $R_{SM}$ for the present data are given in Table 1, where $G_D \equiv 1/(1+Q^2/.71)^2$.
They  are in 
good agreement with the  empirical fit. The resonance
contributions to the amplitudes were also obtained within the framework of the model.
We note that at the resonance position the resonance amplitudes account for nearly
all of $G^\ast_M$, $R_{EM}$, and $R_{SM}$, as indicated in Table 1. These
results for the resonance  are shown in Fig. \ref{results}. The first errors shown
are statistical. The second are  systematic errors 
estimated from variations in the empirical fits of  experimental quantities
resulting from our estimates of  systematic errors in the cross sections. 

\vspace{.15in}

\hspace{-.3in}\begin{minipage}{3.4in}
\begin{flushleft}
{\noindent
\begin{tabular}{|c|c|c|c|}\hline
$Q^2$ &  $G^\ast_M/3G_D$  &  $R_{EM}$ &  $R_{SM}$            \\ \hline\hline
\multicolumn{4}{|c|}{  Resonance + Background}\\ \hline
2.8    & $0.70 \pm .02 \pm .02$  &  $-.023 \pm .012  \pm .005 $ & $ -0.114 \pm .013  \pm .01$ \\ \hline
4.0    & $0.59 \pm .02  \pm .02 $ &  $-.035 \pm .012  \pm .005$  & $ -0.150 \pm .013  \pm .01$  \\ \hline\hline
\multicolumn{4}{|c|}{ Resonance Only} \\ \hline
2.8    & $0.70 \pm .02 \pm .02$& $ -.020 \pm .012  \pm .005$  &  $-0.112 \pm .013  \pm .01$\\ \hline
4.0    & $0.59 \pm .02 \pm .02 $& $ -.031 \pm .012  \pm .005$  &  $ -0.148 \pm .013  \pm $ .01 \\ \hline
\end{tabular}
}

\vspace{.05in}
{Table I. $G^\ast_M$, $R_{EM}$, and $R_{SM}$ extracted from the present data
by means of the effective Lagrangian fits discussed in the text.}
\end{flushleft}
\end{minipage}
\vspace{.15in}

From Fig. 3a, it is seen 
that the  $N-\Delta$ transition
form factor $G^\ast_M$ is decreasing with $Q^2$ faster than the dipole form, which
characterizes the elastic nucleon form factor \cite{bosted} and those for the strongly
excited resonances in the second and third resonance region \cite{sto}.  This result corroborates
earlier fits to the
peaks above the non-resonant continuum in available single arm inclusive  
electron scattering data \cite{sto}\cite{kep}. This $Q^2$ dependence of $G^\ast_M$
is in disagreement
with pQCD constituent scaling predictions. The  $N-\Delta$ transition was previously
studied theoretically in a pQCD framework \cite{carl}, which resulted in an anomalously small 
helicity conserving amplitude compared to that for
elastic and other resonance amplitudes. Thus they conclude that the  pQCD description 
for the $\Delta(1232)$  might become important only at a higher $Q^2$ than for the
elastic or other resonance transitions. 

The dot-dash curve in Figure 3a  is the result  of a relativistic quark
model calculation  \cite{cap}, and clearly does not describe the data. The solid
curve \cite{card}, which more accurately describes the  $Q^2$ dependence of
$G^\ast_M$  is also due to
a relativistic quark model calculation, as in \cite{cap}, but with  a quark 
form factor
added. The dashed curve is the result of a QCD sum-rule calculation \cite{bel},
 which utilizes duality between quark and hadron spectral densities to
parameterize the dominant non-perturbative transition matrix elements. 

Regarding the ratios   $R_{EM}$ and $R_{SM}$,  both are small, as are
the quark model predictions, even taking into account calculational
uncertainties \cite{cap}.  Beyond that  neither quark model calculation
describes  all the data well. On the other hand,  the pQCD prediction 
($R_{EM} = +1$) can be ruled out unambiguously. 

Recently \cite{int} there has been an attempt  to describe  the small value of
$R_{EM}$ in the few GeV$^2$  range of $Q^2$ by  interpolating  between the
pQCD calculated asymptotic amplitudes \cite{carl}, and the 
measured\cite{photo} photoproduction  ($Q^2=0$) amplitude. For the helicity
non-conserving amplitude they used a dipole form with a faster falloff than the
helicity conserving amplitude. The results for $R_{EM}$
remain relatively small at  moderately large momentum transfer, but are still in
significant disagreement with the data.

To  summarize, we  studied  exclusive $\pi^0$ electroproduction in the region of
the 
$\Delta(1232)$ resonance at $Q^2$ = 2.8  and 4.0~GeV$^2$. We find that the ratio 
$R_{EM}$ remains small, the ratio $R_{SM}$ is small but nonzero, and the   $N-\Delta$
transition form factor $G^\ast_M$ is decreasing with $Q^2$ faster than the dipole form factor, 
as suggested by previous analyses of inclusive data.

These results 
indicate that the hadron helicity is not conserved in this  
reaction and that pQCD  is not applicable for this reaction at these 
momentum transfers; on the other hand, it is not clear how high in  $Q^2$
the use of constituent quark models is appropriate.
There has been recent progress  in  treating  exclusive reactions in terms of
{\em off-forward parton distributons}. These  are generalizations of deep inelastic
scattering parton distribution functions,  and include exclusive reactions
and form factors in the $Q^2$ region where soft processes dominate\cite{rad2}.
 The application of these results to the
present reaction, and  to the  higher  $Q^2$  extensions  planned for Jefferson Lab
are anticipated.

The authors  acknowledge the support of the staff of the Accelerator
Division of Thomas Jefferson National Accelerator Facility. This work was supported in part 
by the U.S.~Department of Energy, the National Science Foundation, and the Korean Science 
and Engineering Foundation (KOSEF). We would like to thank S. Capstick,
C. E. Carlson, B. Keister, T.-S. H. Lee,  A. V. Radyushkin,  and S. Simula 
for communicating results of their calculations, as well as for many fruitful discussions. 

$^\ast$ Present address, Physics Department, University of Minnesota, Minneapolis, MN, 55455.

$^+$ Present address, Physics Program, Louisiana Tech University, Ruston, LA 71272.

\end{document}